\documentstyle[epsf,epsfig,rotating,here]{article}

\def\beq{\begin{equation}}
\def\eeq{\end{equation}}
\def\bea{\begin{eqnarray}}
\def\eea{\end{eqnarray}}
\def\bq{\begin{quote}}
\def\eq{\end{quote}}

\def\gappeq{\mathrel{\rlap {\raise.5ex\hbox{$>$}}
{\lower.5ex\hbox{$\sim$}}}}

\def\lappeq{\mathrel{\rlap{\raise.5ex\hbox{$<$}}
{\lower.5ex\hbox{$\sim$}}}}

\begin{document}
\pagestyle{empty}
\begin{flushright}
{CERN-TH/97-367}
\end{flushright}
\vspace*{5mm}
\begin{center}
{\bf PHYSICS AT FUTURE COLLIDERS}
\\
\vspace*{1cm} 
{\bf John ELLIS} \\
\vspace{0.3cm}
Theoretical Physics Division, CERN \\
CH - 1211 Geneva 23 \\
\vspace*{2cm}  
{\bf ABSTRACT} \\ \end{center}
\vspace*{5mm}
\noindent
After a brief review of the Big Issues in particle physics,
we discuss the contributions to resolving that could be made by
various planned and proposed future colliders. These include
future runs of LEP and the Fermilab Tevatron collider, B factories,
RHIC, the LHC, a linear $e^+e^-$ collider, an $e-p$ collider in
the LEP/LHC tunnel, a $\mu^+ \mu^-$ collider and a future larger
hadron collider (FLHC). The Higgs boson and supersymmetry are used as
benchmarks for assessing their capabilities. The LHC has great capacities
for precision measurements
as well as exploration, but also shortcomings where the complementary
strengths of a linear $e^+ e^-$ collider would be invaluable. It is
not too soon to study seriously possible subsequent colliders.

\vspace*{1cm}

\begin{center}
{\it Plenary Session Talk}\\
{\it  presented at the}\\
{\it Europhysics  Conference on High Energy Physics} \\
{\it Jerusalem, August 1997} 
\end{center}
\vspace*{3cm}

\begin{flushleft} CERN-TH/97-367 \\
December 1997
\end{flushleft}

\vfill\eject
\setcounter{page}{1}
\pagestyle{plain}

\section{The Big Issues}

The discoveries for which future colliders will probably be remembered
are not those which are anticipated.  Nevertheless, we cannot avoid
comparing their capabilities to address our present prejudices as to
what big issues they should resolve.  These include, first and foremost,
the {\bf Problem of Mass}: is there an elementary Higgs boson, or is it
replaced by some composite technicolour scenario, and is any Higgs boson
accompanied by a protective bodyguard of supersymmetric particles?  As
we were reminded at this meeting~\cite{Ward}, precision electroweak data
persist in
preferring a relatively light Higgs boson with mass
\beq
m_H = 115^{+116}_{-66} \; {\rm GeV}
\label{one}
\eeq
which is difficult to reconcile with calculable composite scenarios.
However, we should be warned that the range (\ref{one}) is compatible with the
validity of the Standard Model all the way to the Planck
scale~\cite{Planck}, as seen in Fig.~1, %\ref{HR}, 
with no
new physics to stabilize the electroweak coupling or keep the
Standard Model couplings finite.  Nevertheless, the range (\ref{one}) is
highly consistent~\cite{EFL} with the prediction of the minimal
supersymmetric extension of the Standard Model~\cite{MSSMHiggs}, so
we use in this talk
supersymmetry~\cite{susy} as one of our benchmarks for future colliders.
The {\bf
Problem of Flavour} includes the questions why there are just six quarks
and six leptons, what is the origin of their mass ratios and the
generalized Cabibbo mixing angles, and what is the origin of CP
violation? The Standard Model predicts the presence of CP violation, but
we do not yet gave any quantitative tests of the Kobayashi-Maskawa
mechanism~\cite{Jarlskog}: detailed studies may reveal its inadequacy.
Finally, the
{\bf Problem of Unification} raises the possibility of neutrino masses
and proton decay, that are not addressed by colliders.  However, GUTs
also predict many relations between couplings, such as $\sin
^{2}\theta_{W}$~\cite{GQW},
and masses, such as $m_b / m_{\tau}$~\cite{mb}, that can be tested at
colliders, such as detailed predictions for the spectroscopy of
sparticles - if there are any!

\begin{figure}[H]%fig. 1
\hglue5cm
\epsfig{figure=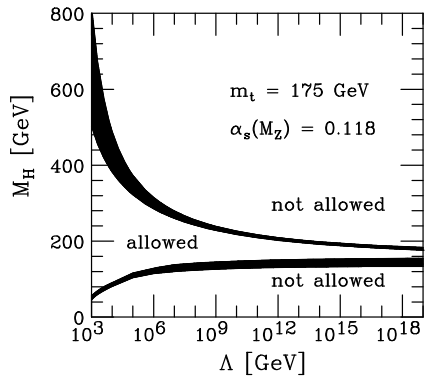,width=5.5cm}
\label{HR}
\caption[]{The range of $M_H$ allowed if the Standard Model remains valid,
unmodified, up to an energy scale $\Lambda$ [2].}
\end{figure}

\section{Catalogue of Future Colliders}

Table 1 lists future high-energy colliders, both approved and under
discussion, together with some of their key parameters and their planned
start-up dates~\cite{Mattig}.  To these should be added LEP 2000, which is
the
proposal to extend the scheduled running of LEP through the year 2000,
for which approval and extra funding is now being sought from the CERN
Member States.  By using some of the LHC cryogenic facilities, this and
the 1999 run of LEP could be at a higher energy, possibly the 200 GeV
foreseen in the original LEP design.  This would extend the LEP reach for
the Standard
Model Higgs boson through the LEP/LHC transition region to above 100
GeV. The principal gain for Higgs physics of the extra year's run would be
to be sure of overcoming the $Z$
background around 90 GeV.  Running LEP at 200 GeV would also extend
significantly the LEP coverage of MSSM Higgs bosons~\cite{Janot}, passing
a
definitive verdict on the region $\tan \beta \lappeq 2$ favoured in many
theoretical models, as seen in Fig.~2, %\ref{MSSM200}, 
and would also
provide closure on supersymmetric
interpretations of the CDF $e^+ e^- \gamma \gamma p_{T}$
event~\cite{CDFgamma}.

\begin{table}[h]
\caption{Table of future collider parameters}
\begin{center}
\begin{tabular}{llllc}
Collider & Particles & $E_{cm}$ (GeV) & Luminosity(cm$^{-2}$s$^{-1}$) & Starting Date \\
&&&&\\
PEP II & $e^+e^-$ & 10 & $3\times 10^{33}$ & 1999 \\
KEK-B   & $e^+e^-$ & 10 & $ 10^{33}$ & 1999 \\
RHIC & $Au-Au$ & 200A & $2\times 10^{26}$ & 1999 \\
LHC & $pp$ & $14\times 10^3$ & $10^{34}$ & 2005 \\
LHC & $Pb - Pb$ & $1.15\times 10^6$ & $10^{27}$ & 2005 \\
LHC & $ep$ & 1300 & $10^{32}$ & ? \\
LC & $e^+e^-$& 500/2000 & $5\times 10^{33}/10^{34}$ & 2008 ?\\
FMC & $\mu^+\mu^-$ & $m_H/4000$ & $2\times 10^{33}/10^{35}$ & ? \\
FLHC & $pp$ & $1/2\times 10^5$ & $10^{34}$ & ?
\end{tabular}
\end{center}
\end{table}

\begin{figure}%fig. 2
\hglue5cm
\epsfig{figure=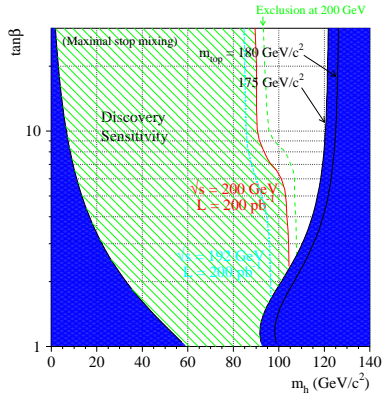,width=5.5cm}
\label{MSSM200}
\caption[]{The sensitivity to MSSM Higgs bosons that may be achieved at LEP
2 [10].}
\end{figure}

Also in the category of future runs of present accelerators is the TeV
2000 programme~\cite{JhButler}.  This actually comprises two
runs: Run II,
which is
approved to run from 1999 to about 2002, and is slated to accumulate $4
fb^{-1}$ of data, and Run III, which is proposed to start
a couple of years later and continue
until the LHC kicks in, gathering about
$20 fb^{-1}$ of data.
Top of the TeV 2000 physics agenda is top
physics~\cite{JhButler}, with such objectives as decreasing the
uncertainty
in $m_t$ to 
$\sim$ 2 GeV in Run II and perhaps 1 GeV in Run III, measuring $\sigma_{\bar tt}$
with an accuracy of 7(5) \% (which would test ``topcolour" models), and
determining $\Gamma_t$ to 20(12) \%. TeV 2000 also offers new prospects in $W$
physics: $\delta m_W \simeq$ 40(20) MeV, $\delta\Gamma \simeq$ 15 MeV and
measuring triple-gauge couplings at the 10 \% level. 
The TeV 2000 programme also has a nice chance of finding the
Higgs boson of the Standard Model, with a reach extending above
100 GeV in Run II and perhaps to 125 GeV in Run III~\cite{JhButler}.
Beyond this Standard Model
physics, TeV 2000 can search for 
gluinos up to 400 GeV or so, and charginos up to about 220 GeV, depending on the
details of the model~\cite{JhButler}.

The next new colliders to start taking data will presumably be the 
$B$ factories at SLAC and KEK, with their
experiments BaBar and Belle~\cite{BaBar}, respectively. Their tasks will be to
understand CP violation, possibly within
the Standard Model by measuring or overconstraining the unitarity
triangle~\cite{Jarlskog}
\beq
V_{ud}V^*_{ub} + V_{cd}V^*_{cb} + V_{td}V^*_{tb} = 0
\label{two}
\eeq
or (we can but hope!) finding new physics beyond it, 
as well as to improve limits
on (and measurements of) rare B decays. 
This will be a crowded field, with HERA-B~\cite{HERAB}
and the Tevatron Run II starting in 1999, as well as BaBar
and Belle. The
field will become even more crowded during the next decade, with
BTeV~\cite{JeButler} perhaps
starting around 2003, as well as CMS~\cite{CMS,CMSB}, ATLAS~\cite{ATLAS}
and
LHC-B~\cite{LHCB} in 2005. To these should be
added CLEO, which will continue to 
provide useful complementary information on B
physics, as well as SLD, which also has
a chance to measure the $B_s$ mixing parameter $x_s$.

Another collider due to start taking data in 1999 is
RHIC, which expects to reach nuclear energy densities $\sim$ 6
GeV/fm$^3$. As we heard here from Plasil~\cite{Plasil}, this will have two
large experiments:
STAR~\cite{STAR} which will measure general event 
charactersitics and statistical signatures,
and PHENIX~\cite{PHENIX} which will concentrate more on hard probes such
as $\ell^+\ell^-$
pairs. RHIC will also have two smaller experiments: PHOBOS~\cite{Wosiek}
and BRAHMS~\cite{Plasil}. In the
longer run, the LHC also offers relativistic heavy-ion collisions at an
nuclear
energy density which is model-dependent, but expected to be considerably higher
than at RHIC. There will be a dedicated heavy-ion experiment at the LHC,
ALICE~\cite{ALICE}, which aims at both statistical signatures and
$\ell^+\ell^-$ pairs. There
is also interest in heavy-ion physics from CMS~\cite{CMS,Kvatadze}, which
may be
able to observe jet
quenching in events with a large$-p_T Z^0$ 
or $\gamma$ trigger, and ATLAS would
also have interesting capabilities for heavy-ion physics.

\section{LHC Physics}

The primary task of the LHC, approved in late 1994 and scheduled for first beams
in 2005~\cite{Gourber}, is to explore the 1 TeV energy range. The two
major ``discovery"
experiments ATLAS~\cite{ATLAS} and CMS~\cite{CMS} were approved in early
1996, and construction of some
of their detectors has begun. ALICE~\cite{ALICE} was approved in early
1997, and the dedicated
CP-violation experiment LHC-B~\cite{LHCB} has passed successfully through
the preliminary
stages of approval and is expected to receive final approval soon.

Top of the physics agenda for the LHC is the elucidation of the origin of
particle masses, i.e., the mechanism of spontaneous electroweak symmetry
breakdown. Within the Standard Model, this means looking for the Higgs boson,
whose mass is currently estimated in (\ref{one}). The branching ratios and
production cross sections for the Higgs at the 
LHC are well understood, including
first-order QCD corrections, as reviewed here by Spira~\cite{Spira} and
shown in Fig.~3. %\ref{spira}.

\begin{figure}%fig. 3
\hspace*{5.5cm}
\begin{turn}{-90}%
\epsfig{figure=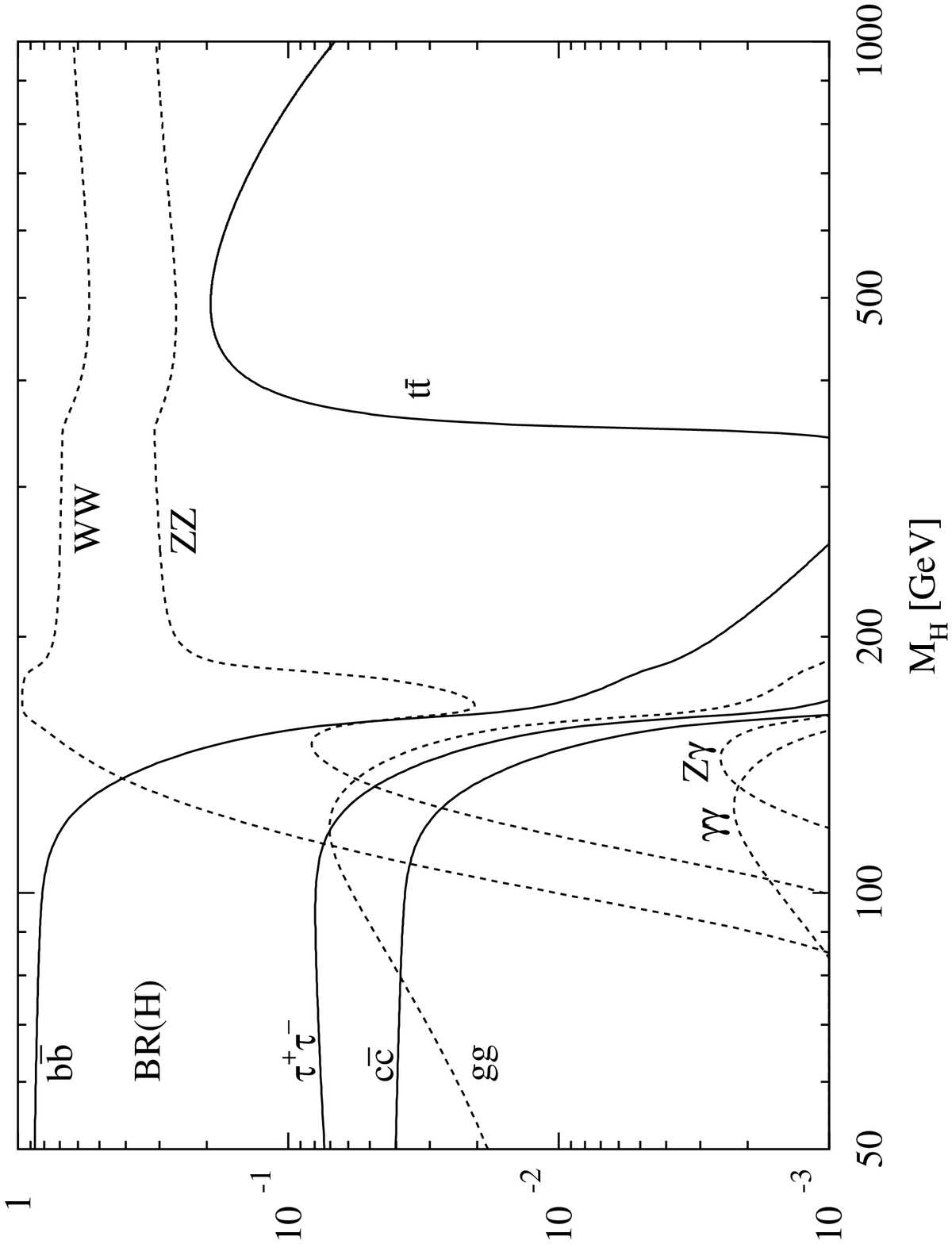,width=3.5cm}
\end{turn} \\[1.0cm]
\hspace*{5.5cm}
\begin{turn}{-90}%
\epsfig{figure=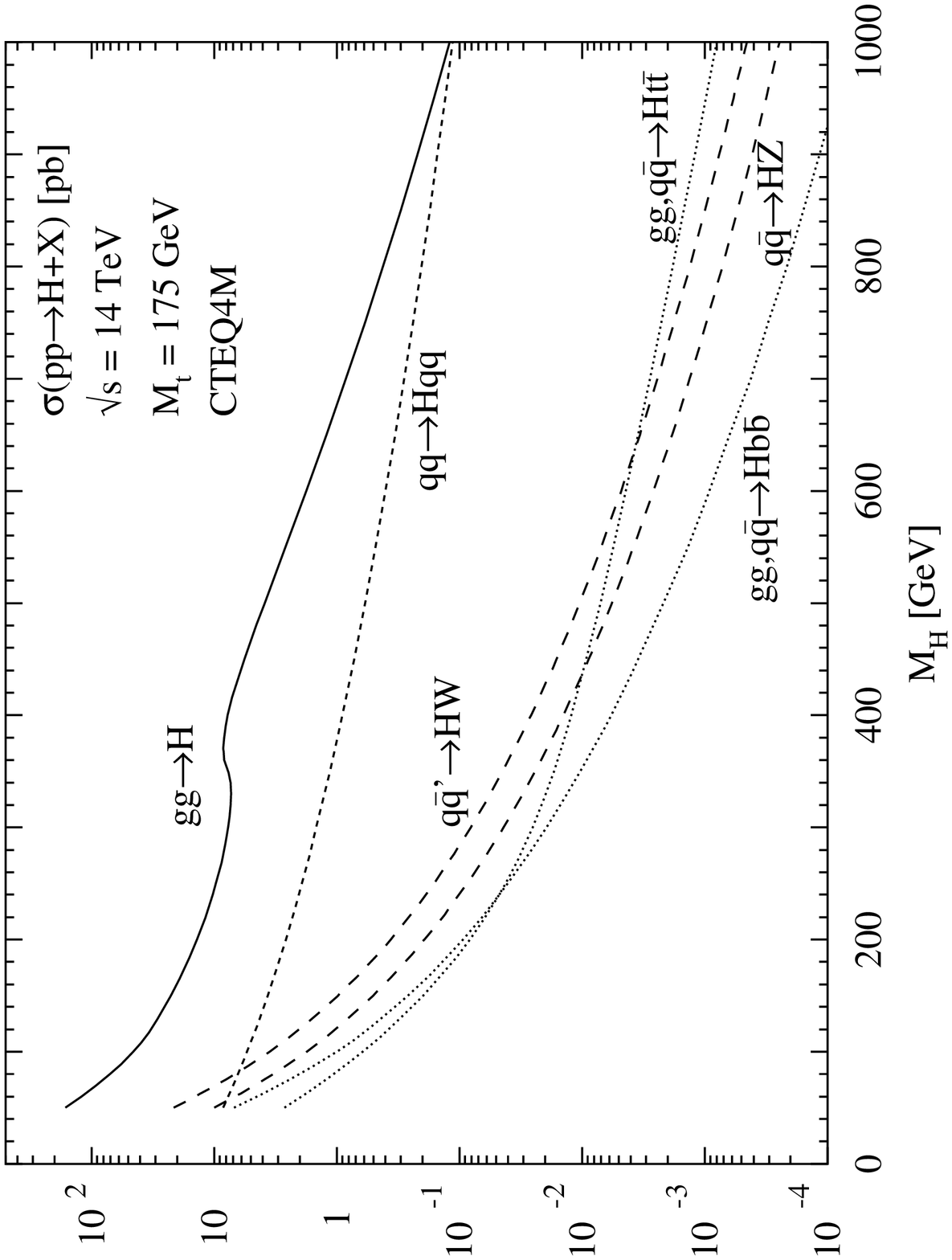,width=3.5cm}
\end{turn}\\[0.5cm]
\label{spira}
\caption[]{ (a) Branching ratios and (b) cross sections for Higgs
production at the LHC, including radiative corrections [28].}
\end{figure}

However, work is still
needed on the QCD corrections to some backgrounds, such as $\gamma\gamma$.
Favoured search signatures~\cite{ATLAS,CMS} include $H
\rightarrow\gamma\gamma$ for 100 GeV
$\lappeq m_H \lappeq$ 140 GeV, $H \rightarrow 4\ell^\pm$ for 130 GeV $\lappeq m_H
\lappeq$ 700 GeV and $H\rightarrow W^+W^-, Z^0Z^0 \rightarrow \ell^+\ell^- jj,
\bar \nu \nu \ell^+\ell^-, \ell\nu jj$ for $m_H \gappeq$ 500 GeV. 
As seen in Fig.~4, %\ref{LHCHiggs}, 
one delicate mass region is $m_H
\lappeq$ 120 GeV, where the requirements on
electromagnetic calorimeters are particularly stringent~\cite{Cashmore}.
Recent studies indicate
that the channel $W + (H\rightarrow \bar bb)$ may play a useful r\^ole in this
mass region~\cite{Poggioli}. Another delicate mass region is $m_H
\sim$ 170 GeV, where the branching ratio for the preferred $H\rightarrow ZZ^*
\rightarrow 4\ell^\pm$ signal is reduced, as seen in
Fig.~4, %\ref{LHCHiggs}, 
because the $H\rightarrow W^+W^-$ channel
opens up. The possibility of isolating the $H\rightarrow 
W^+W^-\rightarrow (\ell^+\nu)(\ell^-\bar\nu)$ decay mode in this mass region has
recently been re-examined~\cite{DD,Poggioli}. By making suitable cuts on
the charged leptons
$(\vert\eta\vert < 2.4, p_T >$ 25 GeV, 
$p_{T_2} >$ 10 GeV, $m_{\ell^+\ell^-} >$
10 GeV, $\vert m_{\ell^+\ell^-}-m_Z\vert >$ 5 GeV) and vetoing events with jets
(no $p_{T_j} >$ 20 GeV in $\vert\eta\vert < 3$), 
as well as cuts on the polar and
transverse opening angles of the $\ell^+\ell^-$, it was found
possible to display
an excess above the continuuum $W^+W^-$ background, which is
clearest when $m_H
\sim$ 170 GeV, as seen in Fig.~5. %\ref{michaelboth}.
 However, this technique does
not produce a well-defined resonance peak.

\begin{figure}%fig. 4
\hglue5cm
\epsfig{figure=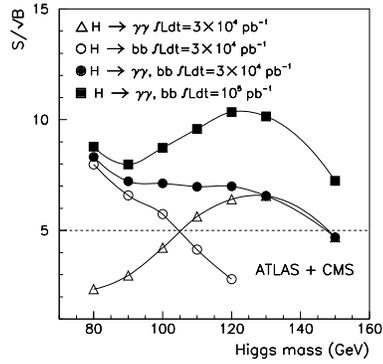,width=5.5cm}
\label{LHCHiggs}
\caption[]{Estimated significance of light-mass Higgs detection at the LHC:
note the ``delicate" regions $m_H \lappeq$ 120 GeV and $\sim$ 170 GeV [30].}
\end{figure}

\begin{figure}%fig. 5
\hglue5cm
\epsfig{figure=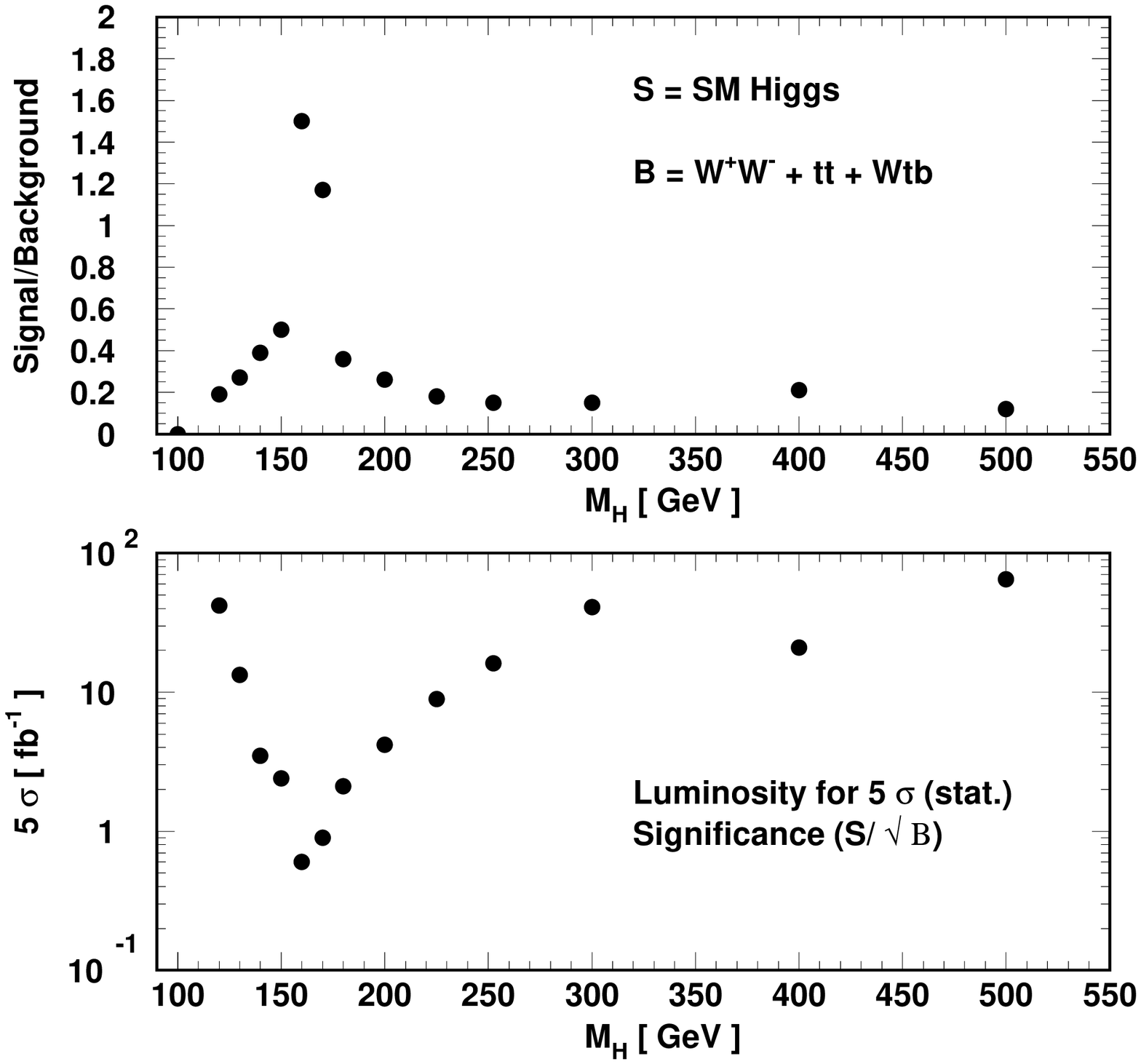,width=5.5cm}
\label{michaelboth}
\caption[]{Possible improvement in the sensitivity to $m_H \sim$ 170 GeV at
the LHC, using cuts optimized for $H\rightarrow W^+W^-$ decay [31].}
\end{figure}

It is hard to find any theorist who thinks that the single Higgs boson of the
Standard Model is the whole of the story, and many would plump for the minimal
supersymmetric extension of the Standard Model (MSSM). The mass of the lightest
neutral Higgs boson $h$ in the MSSM is restricted to $m_h \lappeq$ 120
GeV~\cite{MSSMHiggs}, which
is, as already mentioned, quite consistent~\cite{EFL} with the range
(\ref{one})
indicated by precision electroweak
measurements~\cite{Ward}. The MSSM Higgs branching ratios and production
cross sections at
the LHC have also been studied intensively, including leading-order QCD
corrections~\cite{Spira}. It has been known for some time that, although
there are extensive
regions of the MSSM Higgs parameter space where one or more of the MSSM Higgs
bosons may be detected~\cite{Tuominiemi}, there is a region around $m_A
\sim$ 150 to 200 GeV and
$\tan\beta \sim$ 5 to 10 which is difficult to cover. As seen in
Fig.~6, %\ref{elzbieta}, 
after
several years of running the LHC at the design luminosity, even this region may
be covered by combining data from ATLAS and CMS~\cite{Tuominiemi}.
However, one would be more
comfortable if there were more coverage, and in 
particular with more help from LEP by covering a
larger area of the $(m_A, \tan\beta)$ plane by running at $E_{cm}$ = 200
GeV~\cite{Janot}.

\begin{figure}%fig. 6
\hglue5cm
\epsfig{figure=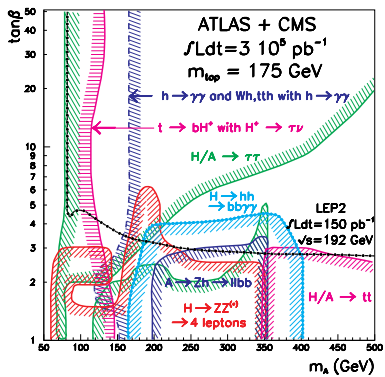,width=5.5cm}
\label{elzbieta}
\caption[]{Detectability of MSSM Higgs bosons at the LHC. In most of the
plane, more than one experimental signature is visible [32].}
\end{figure}

We can still hope that supersymmetry may be discovered before the start-up
of the LHC~\cite{Dittmar}, but
the LHC has unprecedented mass reach in the search for
supersymmetric particles~\cite{Paige}. Cross sections for producing the
strongly-interacting
squarks $\tilde q$ and gluons $\tilde g$ have been calculated including
leading-order QCD corrections~\cite{Zerwas}. Since these are expected to
be among the heaviest
sparticles, if $R$ parity is conserved one expects their generic decays to
involve complicated cascades such as $\tilde g \rightarrow \tilde b \bar b,
\tilde b \rightarrow \chi_2 b, \chi_2 \rightarrow \chi_1 \ell^+\ell^-$ where
neutralinos are denoted by $\chi_i$. Therefore generic signatures are missing
energy, leptons and hadronic jets (which may include $b$
quarks)~\cite{Abdullin}. These are also
interesting signatures if $R$ parity is violated~\cite{Baer}, with the
added possibility of
reconstructing mass bumps in lepton + jet combinations. As is well known, the
$R$-conserving missing-energy signal would stick out clearly above the Standard
Model background, enabling $\tilde q$ and 
$\tilde g$ with masses between about 300
GeV and 2 TeV to be discovered, as discussed here by~\cite{Paige}.

The main recent novelty has been the realization that the $\tilde q$ or $\tilde g$
decay cascades may be reconstructed and detailed spectroscopic
measurements made~\cite{HSPSY}
The following is the basic strategy proposed.
\begin{enumerate}
\item Identify a general supersymmetric signal, e.g., in four-jet + missing
energy events via the global variable
\beq
m_{eff} = p_{T_{j_1}} + p_{T_{j_2}} + p_{T_{j_3}} + p_{T_{j_4}} + 
p\llap{$/$}_t
\label{three}
\eeq
as seen in Fig.~7, %\ref{susymeff}, 
whose mean is found from the Monte Carlo
studies to be around $2m_{(\tilde q~{\rm
or}~\tilde g)}$.
\item Reconstruct decay chains starting from the end, e.g., in the above case via
$\chi_2\rightarrow \chi_1 (\ell^+\ell^-)$ which should exhibit a characteristic
edge in the spectrum at $m_{\ell^+\ell^-} = m_{\chi_2} -
m_{\chi_1}$, then adding (for example)
$b$ and $\bar b$ jets to reconstruct $m_{\tilde b}$ and $m_{\tilde g}$.
\item Finally,  make a global fit to MSSM parameters within an assumed standard
parametrization such as that suggested by supergravity, namely $m_{1/2}$ (a
common gaugino mass), $m_0$ (a common scalar mass), $A$ (a common trilinear soft
supersymmetry-breaking parameter), $\tan\beta$ (the ratio of Higgs vacuum
expectation values), and the sign of the Higgs mixing parameter
$\mu$~\cite{susy}.
\end{enumerate}

\begin{figure}%fig. 7
\hglue5cm
\epsfig{figure=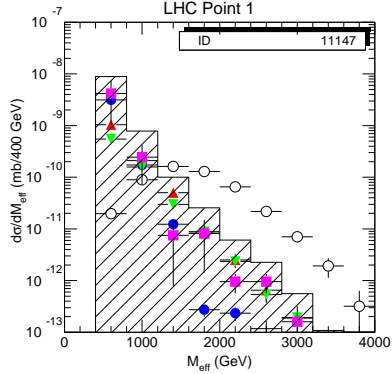,width=5.5cm}
\label{susymeff}
\caption[]{Significance of $m_{eff}$ (3) signal at the LHC, compared to
Standard Model backgrounds [38].}
\end{figure}

This strategy has been applied within the framework of a study commissioned by
the LHC experiments committee of five particular points in this parameter
space~\cite{LHCCsusy}.
The values of representative sparticle masses for these parameter choices are
shown in Table 2. A typical $(\ell^+\ell^-)$ 
spectrum is shown in Fig.~8, %\ref{susyedge}, 
where
we see a sharp edge that would enable $m_{\chi_2}-m_{\chi_1}$, to be measured
with a precision of 100 MeV~\cite{Paige,HSPSY}. Events close to this edge
can be used to reconstruct
$\tilde q\rightarrow q\chi_2$ decays, such as
$\tilde b\rightarrow b\chi_2$ and $\tilde g\rightarrow\tilde b\bar b$
decays~\cite{Paige,HSPSY}, as seen in
Fig.~9. %\ref{sbottom}.
For generic other parameter choices, as seen in Fig.~10, %\ref{cascadeh},
one may
reconstruct $h\rightarrow \bar bb$ decays in the
cascade~\cite{Paige,Abdullin}. Another possibility
discussed here~\cite{Paige} is that $\chi_2\rightarrow\chi_1 +
(\tau^+\tau^-)$
decays dominate at large $\tan\beta$, 
in which case one may observe an excess in
the $M_{\tau^+\tau^-}$ distribution. 
Table 3 shows the MSSM particles that may be
discovered at each of the five points in parameter space that have been explored
in detail~\cite{LHCCsusy}. We see that a sizeable fraction of the spectrum
may be accessible at the LHC.

\begin{table}[h]
\caption{Test points for supersymmetry studies at the LHC (masses in GeV)}
\begin{center}
\begin{tabular}{cllcrrlrlr}
 & $m_0$ & $m_{1/2}$ & $A_0$ & tan$\beta$ & $m_{\tilde g}$ & $m_{\tilde u_R}$ &
$m_{\chi^\pm}$ & $m_{\tilde e_R}$ & $m_h$ \\
&&&&&&&&&\\
1 & 400 & 400 & 0 & 2 & 1004 & 925 & 325 & 430 & 111 \\
2 & 400 & 400 & 0 & 10 & 1008 & 933 & 321 & 431 & 125 \\
3 & 200 & 100 & 0 & 2 & 298 & 313 & 96 & 207 & 68 \\
4 & 800 & 200 & 0 & 10 & 582 & 910 & 147 & 805 & 117 \\
5 & 100 & 300 & 300 & 2.1 & 767 & 664 & 232 & 157 & 104
\end{tabular}
\end{center}
\end{table}

\begin{table}[h]
\caption{The LHC as ``Bevatrino":Sparticles detectable at selected points in
supersymmetric parameter space and denoted by +}
\begin{center}
\begin{tabular}{ccccccccccccc}
 & $h$ & $H/A$ & $\chi^0_2$ & $\chi^0_3$ & $\chi^-_1$  &$\chi^\pm_1$ & $\chi^\pm_2$ &
$\tilde q$ & $\tilde b$ & $\tilde t$ & $\tilde g$ & $\tilde\ell$ \\
&&&&&&&&&&&& \\
1 & + && + &&&&& + & + & + & + &\\
2 & + &&+ &&&&& + & + & + & + & \\
3 & + & + & + &&&+ && + & + && + & \\
4 & + && + & + & + & + & + & + &&& + & \\
5 & + && + &&&&& + & + & + & + & +
\end{tabular}
\end{center}
\end{table}

\begin{figure}%fig. 8
\hglue5cm
\epsfig{figure=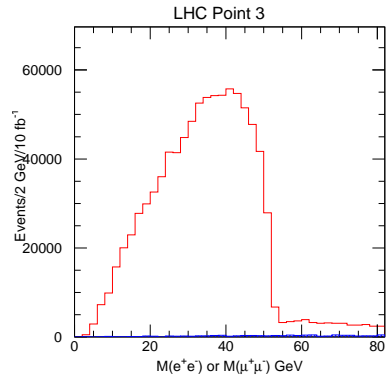,width=5.5cm}
\label{susyedge}
\caption[]{Typical $\ell^+\ell^-$ spectrum from $\chi_2\rightarrow
\chi_1,\ell^+\ell^-$ decay in cascade decays of $\tilde q/\tilde g$ at the
LHC [38].}
\end{figure}

\begin{figure}%fig. 9
\hglue5cm
\epsfig{figure=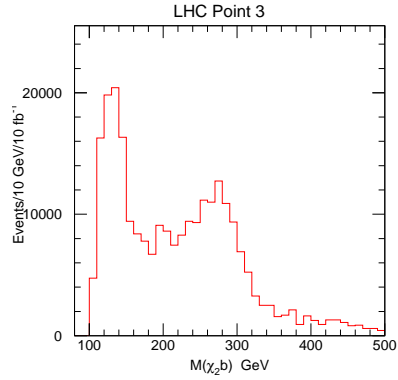,width=5.5cm}
\label{sbottom}
\caption[]{Typical $\tilde b\rightarrow \chi_2 b$ decay signature at the LHC
[38].}
\end{figure}

\begin{figure}%fig. 10
\hglue5cm
\epsfig{figure=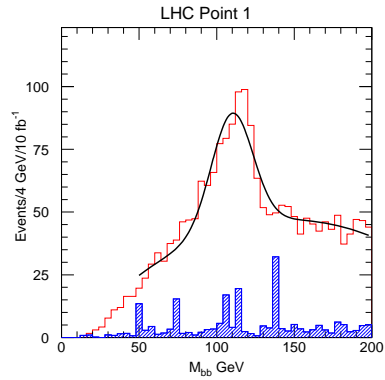,width=5.5cm}
\label{susymeff}
\caption[]{Typical $h\rightarrow \bar bb$ decay signal from cascade $\tilde
q/\tilde g$ decays at the LHC [38].}
\end{figure}

Moreover, precision determinations of the supergravity model parameters are
possible if one combines the different measurements of endpoints, masses,
products of cross sections and branching ratios, etc.~\cite{Paige,HSPSY}. 
For example, it has been estimated that one could attain
\beq
\Delta (m_{\chi_2}-m_{\chi_1}) = \pm \left\{\matrix{50~{\rm MeV} \cr ~1~{\rm
GeV}}\right. @~ {\rm point}~\left\{\matrix{3\cr 4}\right.
\label{four}
\eeq
\beq
\matrix{
&\Delta m_{\tilde b_1} = \pm 1.5 \Delta m_{\chi_1} \pm 3~{\rm GeV} \cr
&\Delta (m_{\tilde g}-m_{\tilde b_1}) = \pm 2~{\rm GeV}\hfill} \left\}
\matrix{@~{\rm point}~3 \cr }\right.
\label{five}
\eeq
A global fit at point 5 yielded
\beq
\Delta m_0 = \pm 5(\pm 3)~{\rm GeV},~~
\Delta m_{1/2} = \pm 8(\pm
4)~{\rm GeV}~,
\nonumber~~
\Delta \tan\beta = \pm 0.11 (\pm 0.02) 
\label{six}
\eeq
where the different errors refer to different stages in the sophistication of the
analysis, and other examples are shown in~\cite{Paige}.

These are not the only precision aspects of physics at the LHC. Parton
distributions are the limiting uncertainties in present experiments, cf the
large-$E_T$ jet cross section at FNAL~\cite{largeet} and 
the large-$x$ data at HERA~\cite{HERA}. Previously,
it had been thought difficult to determine the $pp$ collision luminosity with a
precision better than 5\%. A new approach~\cite{Behner,BDPZ} is to bypass
this step, and measure
directly the parton-parton luminosity functions via the rapidity distributions
of the $W^\pm, Z^0$, which could fix the products $q(x_1)\bar q(x_2)$ for
10$^{-4} \gappeq x \gappeq$ 0.1 with an accuracy of $\pm 1/2 \%$, cf the LEP
luminosity error of $\pm 1/4 \%$ from Bhabha scattering. The problem in
exploiting this is primarily theoretical: can one relate the cross section
for
$W^\pm$ or $Z^0$ production to other production cross sections, e.g., for
$W^+W^-$ pairs or the Higgs, with comparable accuracy? This would require a
significant advance in the state of the art of higher-order QCD calculations.
So far, one-loop corrections to jet cross sections are known~\cite{ES},
including those
due to strongly-interacting sparticle loops~\cite{ER}. These amount to
several percent in
some subprocess cross sections, and may have observable structures at to
the
supersymmetric threshold $M_{jj} = 2m_{\tilde q}$ or $m_{\tilde q} + m_{\tilde
g}$ or $2m_{\tilde g}$.

Beyond the approved (or almost approved) parts of the LHC programme, there are
two further experimental initiatives now circulating. One is for a
full-acceptance $(\vert\eta\vert < 11)$ detector called FELIX~\cite{FELIX} 
to operate at
moderate luminosity $({\cal L}_{pp} \gappeq 2\times 10^{32}$ cm$^{-2}s^{-1})$
with a physics agenda centred on QCD. This would include novel hard phenomena,
such as small-$x$ physics, had diffraction, rapidity-gap physics, the BFKL
pomeron and minijets, as well as soft QCD effects at large impact parameters
and in central collisions. The proposed programme also includes hard and soft
QCD in $pA$ collisions and $\gamma\gamma$ physics in $pp$ and $AA$ collisions,
as well as the search for an astroparticle connection to ``anomalies" in
cosmic-ray collisions, possibly via disordered chiral condensates. FELIX would
require bringing the LHC beams into collision in a new interaction point,
and
has yet to leap the hurdle of assembling a large involved community of
experimentalists.

TOTEM~\cite{TOTEM} is a modest ``classical" proposal to measure the total
$pp$ cross section
$(\Delta\sigma \sim$ 1 mb), elastic scattering in the range $5\times 10^{-4}$
GeV$^2 \gappeq \vert t\vert \gappeq$ 10 GeV$^2$ and diffractive production of
systems weighing up to 3 TeV. It could very likely be placed at the same
interaction point as one of the (almost) approved experiments.

To conclude this review of the LHC physics programme, and set the stage for
proposed subsequent accelerators it is appropriate to conclude this section 
by reviewing some potential sweakenesses of the LHC, as regards probing the
MSSM. It is not suitable for producing and studying the heavier charginos
$\chi^\pm_2$ and neutralinos $\chi_{3,4}$, nor sleptons if they weigh
$\gappeq$
400 GeV. It may have difficulty studying the heavier MSSM Higgs bosons $H, A,
H^\pm$, and even the lightest Higgs boson $h$ if it has non-standard decay
modes. It is not well suited for measuring squark mass differences, since,
e.g., it cannot distinguish $\tilde u, \tilde d$ and $\tilde s$. Thus it is
questionable whether it provides enough cross-checks on the validity of the
MSSM and test simplified parametrizations such as those based on supergravity
with a universal scalar mass $m_0$. There will be plenty
of scope for
further studies of supersymmetry, even if it discovered earlier at the LHC (or
LEP or the FNAL collider).

\section{$e^+e^-$ linear collider physics}

These machines offer a very clean experimental environment and egalitarian
production of new weakly-interacting particles. Moreover, polarizing the beam
is easy and can yield interesting physics signatures, and $e\gamma,
\gamma\gamma$ and $e^-e^-$ collisions can also be arranged quite easily. Thus
linear colliders have many features complementary to those of the
LHC~\cite{Wiik}.

It is likely that the first linear collider would have an initial
centre-of-mass energy of a few hundred GeV. Some of the cross sections for
important physics processes~\cite{Martinez} are shown in
Fig.~11~\cite{ECFA}. %\ref{sigma}
In the absence (so far) of
any clear indication of a threshold in this energy range for new physics, we
start by reviewing the bread-and-butter Standard Model physics agenda of such a
machine. Pride of place goes to top physics. Detailed measurements of
$\sigma_{\bar tt}$ and momentum spectra around the threshold at $E_{cm} \sim$
350 GeV, shown in Fig.~12, %\ref{topthresh}, 
should enable the error in the
top quark mass to be reduced to $\sim$
120 MeV, and it should be possible to measure $\Gamma_t$ with an error around
10\%. It will also be possible to search very cleanly for non-standard top
decays and measure static parameters of the $t$ quark, such as $g^Z_{A,V}$,
$\mu_t$, its Higgs coupling~\cite{BarSh} and its electric dipole moment.
Turning to $W^\pm$ and $Z^0$ physics,
precision on the triple-gauge couplings can be improved over the LHC down to
the 10$^{-3}$ level. Moreover, if one is able to run the collider with high
luminosity at the $Z^0$ peak, one can quickly   obtain a precise measurement of
$\sin^2\theta_W$ ($\pm$ 0.0001), and running at the $W^+W^-$ threshold could
enable the error on $m_W$ to be reduced to 15 MeV.

\begin{figure}%fig. 11
\hglue5cm
\epsfig{figure=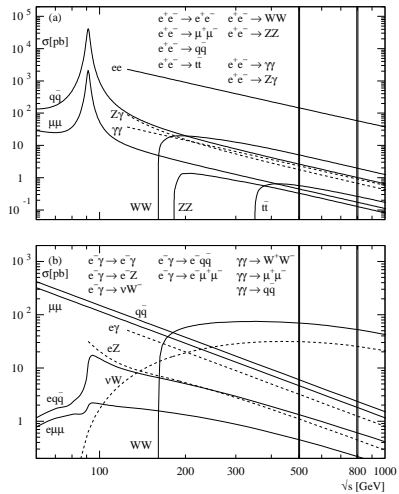,width=5.5cm}
\label{sigma}
\caption[]{Important cross sections at a linear collider [50].}
\end{figure}

\begin{figure}%fig. 12
\hglue5cm
\epsfig{figure=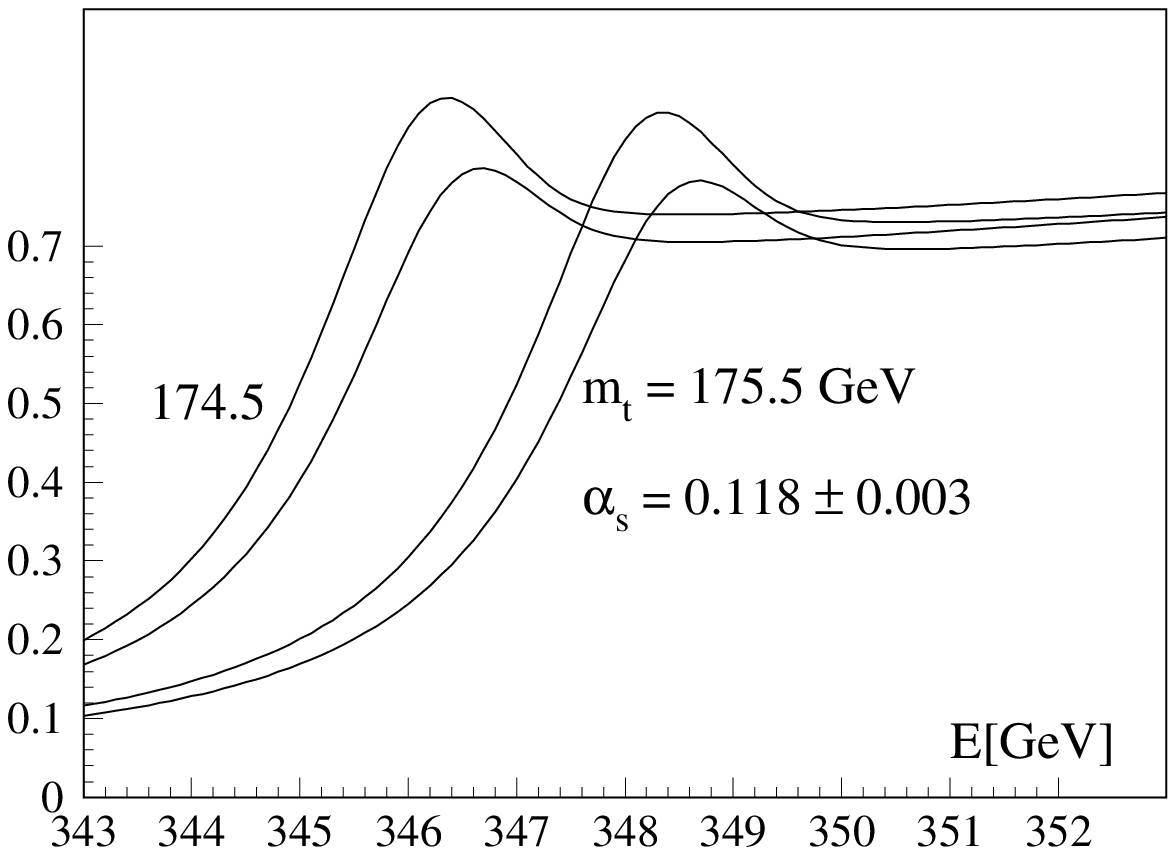,width=5cm}\\
\hglue5cm
\epsfig{figure=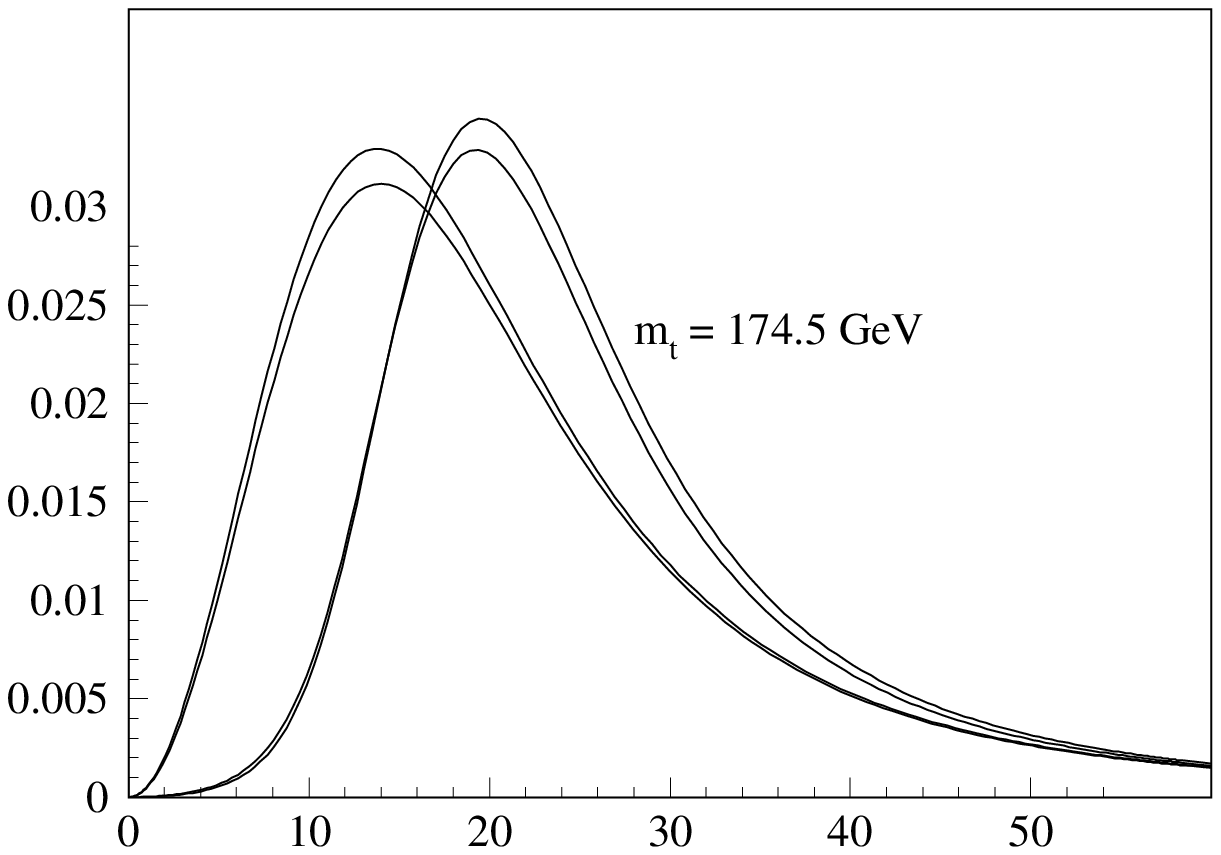,width=5cm}
\label{topthresh}
\caption[]{(a) Cross section and (b) kinematic measurements of
$e^+e^-\rightarrow \bar tt$ at a linear collider [50].}
\end{figure}

The range of Higgs masses preferred (\ref{one}) by the precision electroweak
measurements gives hope that the Higgs boson may lie within the kinematic reach
of a first linear collider, via the reactions $e^+e^-\rightarrow Z+H$ and
$e^+e^-\rightarrow H\bar\nu\nu$. Moreover, it is easy to detect a Higgs
boson in the mass ranges that are ``delicate"
at the LHC, namely $m_Z \gappeq m_H \gappeq$ 120 GeV and $m_H \sim$ 170
GeV~\cite{ECFA,Schreiber}.
Even if the Higgs discovery is made elsewhere, a linear collider could tell us
much more about its couplings and branching ratios: $g_{ZZH}, B(\bar bb),
B(WW^*), B(\tau^+\tau^-)$ and $B(\bar cc + gg)$, 
as seen in Fig.~13, %\ref{HiggsBR}, 
enabling us to verify that it
does its job of giving masses to ghe gauge bosons, quarks and leptons, and
giving us a window on possible non-minimal Higgs models. One can also measure
$\Gamma(\gamma\gamma)$ using the $\gamma\gamma$ collider modes and the
spin-parity of the Higgs can be measured~\cite{ECFA}. Turning to the MSSM,
production and
detection of the lightest MSSM Higgs $h$ is guaranteed, and the heavier Higgs
bosons $H^\pm, H, A$ can also be observed if the beam energy is high enough.

\begin{figure}%fig. 13
\hglue5cm
\epsfig{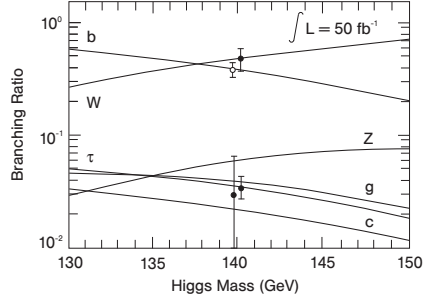}
\label{HiggsBR}
\caption[]{Accuracy with which Higgs decay branching ratios may be measured at a
linear collider [50].}
\end{figure}

As for supersymmetry proper, if its beam energy is above threshold, a linear
collider will produce cleanly electroweakly-interacting sparticles such as the
$\tilde\ell^\pm, \tilde\nu, \chi^\pm$ and $\chi_i$ that are problematic at the
LHC~\cite{ECFA,Martyn}, as seen in Fig. 14. Moreover, sparticle masses can 
be measured
accurately:
\bea
&&\delta m_{\tilde\mu}\sim 1.8~{\rm GeV}, \delta m_{\tilde\nu} \sim 5 ~{\rm GeV},
\nonumber \\
&&\delta m_{\chi^\pm} \sim 0.1~{\rm GeV}
\nonumber \\
&&\delta m_\chi \sim 0.6~{\rm GeV}, \delta m_{\tilde t}\sim 4~{\rm GeV}
\label{seven}
\eea 
enabling one to test supergravity mass
relations, over-constrain model
parameters and check universality (is $m_{\tilde \ell} = m_{\tilde\mu} =
m_{\tilde\tau}$, for example?). Moreover, the couplings and spin-parities of
manuy sparticles can be measured. Thus a linear collider will certainly be able
to add significantly to our knowledge of supersymmetry~\cite{ECFA}, even
if the LHC
discovers it first, and despite the large range of measurements possible at the
LHC -- provided the linear collider beam energy is large enough!

\begin{figure}%fig. 14
\hglue5cm
\epsfig{figure=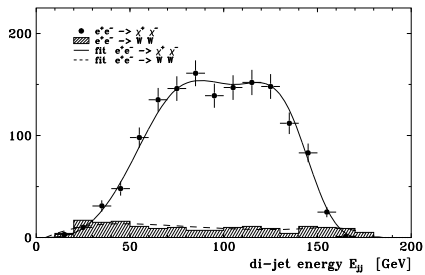,width=6.5cm}
\label{mchargino}
\caption[]{Possible measurement of $e^+e^-\rightarrow \chi^+\chi^-,
\chi^\pm\rightarrow jj\chi$ at a linear collider [50].}
\end{figure}

In my view, we will need a linear collider to complete our exploration of the
TeV energy range, begun by the LHC, to pin down the mechanism of electroweak
symmetry breaking, and to complement the LHC programme with more precision
measurements. To have physics reach comparable to the LHC, the collider energy
should be able to reach $E_{cm} \sim$ 2 TeV, and it would be desirable to be
able to operate back in the LEP energy range and the $Z$ peak and the $W^+W^-$
threshold. Thus the machine should be flexible, with an initial $E_{cm}$ in the
few hundred GeV and the possibility of subsequent upgrades. Unfortunately, we
do not yet know exactly where to start, in the absence of clear information on
new physics thresholds. A final personal comment is that I hope very much that
the linear collider community can converge on a single project. Can the world
afford two such colliders? In the recent past, our political masters have
decided to support just one hadron collider.

\section{Beyond the Standard Colliders}

Even though construction of the LHC will not be completed for another 8 years,
and no linear collider proposal has even been submitted, it is already time to
think what might come next, since we need to maintain a long-term vision of the
future of accelerator-based particle physics, and the $R\times D$ lead time for
any new accelerator project is necessarily very long.

One possible future option which as been kept in mind since the inception of
the LHC project has been an ep collider in the LEP tunnel, using an LHC beam
and an $e^\pm$ beam circulating in a rearranged LEP ring~\cite{LEPLHC}.
The latest design
envisages beam energies of 7 TeV and 67 GeV, yielding collisions at a
centre-of-mass energy of 1.37 TeV with a luminosity ${\cal L}_{ep} \sim
10^{32} cm^{-2}s^{-1}$. The physics interest of this ep option will be easier
to judge after the physics potential of the HERA collider (and in particular
the interpretation of the current large-$Q^2$ anomaly~\cite{HERA}) has
been further
explored: certainly it would be great for producting leptoquarks or
$R$-violating squarks up to masses around 1 TeV.

However, this option is presumably not a complete future for CERN, let alone
the world high-energy physics community. More complete possibilities for the
future are $\mu^+\mu^-$ colliders~\cite{mumu}, which may be the
best way to
collide leptons at $E_{cm} =$ 4 TeV or more, and a possible next-generation
$pp$ collider (variously named the Eloisatron, RLHC or VLHC, called here a
`Future Larger
Hadronic Collider' or FLHC) with $E_{cm}$ = 100 to 200
TeV~\cite{FLHC}. We now
discuss each of these possibilities in turn.

Many technical issues need to be resolved before a multi-TeV $\mu^+\mu^-$
collider can be proposed: the accumulation of the $\mu^\pm$, their cooling,
shielding the detectors and the surrounding populace from their decay
radiation, etc. These should be addressed by a smaller-scale demonstrator
project, much as the SLC demonstrates the linear-collider principle. A very
exciting possibility for this demonstration is a Higgs
factory~\cite{Gunion}, exploiting the
non-universality of the $HL^+L^-$ couplings, which implies that
$\sigma(\mu^+\mu^-\rightarrow H)/\sigma(e^+e^-\rightarrow H) \sim$ 40,000, and
the possibly reduced energy spread in $\mu^+\mu^-$ collisions, which may be as
small as 0.01\%. Neglecting the energy spread, the $\mu^+\mu^-$ cross 
section
in the neighbourhood of the $H$ peak is given by 
\beq
\sigma_H(s) = {4\pi~\Gamma(H\rightarrow \mu^+\mu^-)~\Gamma(H\rightarrow X)\over
(s-m^2_H)^2 + m^2_H \Gamma^2_H}
\label{eight}
\eeq
where the natural width of a 100 GeV Higgs is expected to be about 3 MeV,
whereas $\Delta\sqrt{s}$ may be as low as 10 MeV. Typical line shapes for
Standard Model and MSSM Higgs bosons in this mass range are shown in
Fig.~15. %{mumuHiggs}.
Such a Higgs factory would be able to measure Higgs decay branching ratios into
channels such as $\bar bb, \tau^+\tau^-$, $WW^*$ and $ZZ^*$. It could also draw
a clear distinction between the Standard Model $H$ and the MSSM $h$, could (at
higher energies) separate the $H$ and $A$ of the MSSM, and also make detailed
studies of their properties. Other possible applications of the narrow
$\mu^+\mu^-$ spread in 
$E_{cm}$ include the measurement of $m_H$ with a precision $\sim$ 45 MeV, and
improved precision in the values of $m_t$ and $m_W$~\cite{mumu}.

\begin{figure}%fig. 15
\hglue5cm
\epsfig{figure=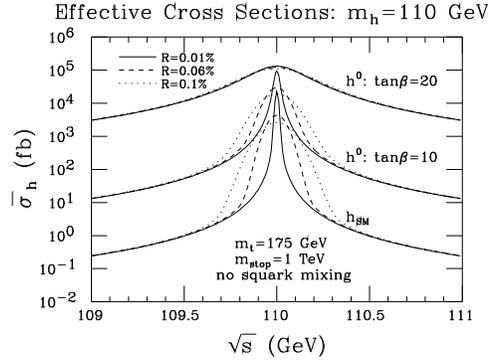,width=6.5cm}
\label{mumuHiggs}
\caption[]{Cross sections for Higgs production at a $\mu^+\mu^-$ collider run as
a ``Higgs factory" [55,57].}
\end{figure}

As for a possible FLHC, this is clearly the tool of choice for exploring
the 10
TeV energy range, as is shown by one example in
Fig.~16~\cite{MM}. %\ref{FLHC}
 At
present, we do not
know what physics may lie there -- the messenger sector of a gauge-mediated
scenario for supersymmetry breaking~\cite{susy}? a
$Z^\prime$ ? a fifth dimension? We will
never know unless we go and look.

\begin{figure}%fig. 16
\hglue5cm
\epsfig{figure=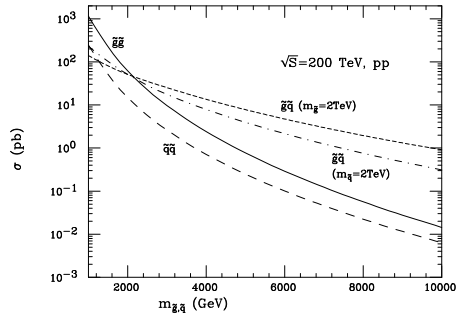,width=6.5cm}
\label{FLHC}
\caption[]{Cross sections for $\tilde q$ and $\tilde g$ production at a FLHC
[58].}
\end{figure}

\vfill\eject

%%%%%%%%%%%%%%%%%%%%%%%%%%
%As shown in reference \cite{ref}, this reads:
%\begin{equation}
%m(r)=\int^r_04\pi r^2 \rho \,dr  \enspace .\label{mass}
%\end{equation}
%Here we reference to equation \ref{mass}

% ---- Bibliography ----
%

\end{document}